\documentclass[runningheads]{llncs}
\usepackage{graphicx}
\usepackage{times}
\usepackage[hidelinks]{hyperref}
\usepackage{xcolor}
\usepackage{cite}

\begin{document}
\title{Collaborative filtering to capture AI user's preferences as norms\thanks{Research funded by the project SAIS: Secure AI AssistantS via Grant EP/T026723/1 from the UK Engineering and Physical Sciences Research Council.}}
%
%
\author{Marc Serramia\inst{1} \and Natalia Criado\inst{2} \and Michael Luck\inst{1}}
\authorrunning{M. Serramia, N. Criado, and M. Luck}
%
\institute{Department of Informatics, King's College London \and Escuela Técnica Superior de Ingeniería Informática, Universitat Politècnica de València}
\maketitle              
\begin{abstract}
Customising AI technologies to each user's preferences is fundamental to them  functioning well. Unfortunately, current methods require too much user involvement and fail to capture their true preferences. In fact, to avoid the nuisance of manually setting preferences, users usually accept the default settings even if these do not conform to their true preferences. Norms can be useful to regulate behaviour and ensure it adheres to user preferences but, while the literature has thoroughly studied norms, most proposals take a formal perspective. Indeed, while there has been some research on constructing norms to capture a user's privacy preferences, these methods rely on domain knowledge which, in the case of AI technologies, is difficult to obtain and maintain. We argue that a new perspective is required when constructing norms, which is to exploit the large amount of preference information readily available from whole systems of users. Inspired by recommender systems, we believe that collaborative filtering can offer a suitable approach to identifying a user's norm preferences without excessive user involvement.
\end{abstract}
\section{Introduction}
\label{sec:intro}

Artificial Intelligence (AI) technologies are becoming commonplace in our lives. From smart watches to AI assistants they bring new functionalities to help us with daily tasks. However, each user of these technologies expects them to function in different ways. For example, in the case of AI assistants, different users have different privacy preferences and would expect the AI assistants to adhere to them. Some might prefer to share their data to enable a tailored experience, while others might prefer to keep their data private. While these AI devices and services usually allow users to customise their preferences, it is far too common that users disregard this option and accept the default settings. For example, a large proportion of users in social networks do not change default privacy settings \cite{Krishnamurthy2009leakage}, demonstrating a wider problem, which is that while users expect AI devices and services to act as they desire, they don't want to invest time in customising them. A common approach to capturing user preferences is to question users when necessary. Unfortunately, the effects of this approach can be even worse, as receiving constant questioning both annoys users and triggers automatic acceptance of the default settings, as mentioned above. To address this problem we must take a smarter approach that does not rely on excessive user involvement.

In order to solve this problem of customising AI devices and services without excessive user input, we aim to exploit a common aspect of these technologies. In particular, we can see these technologies not as individual devices or services of an individual user, but as a multiagent system. 
Here, each user is represented by an agent (be it a smart watch, an AI assistant, an app, a service, etc.) and these agents interact with other agents (other devices, third party services or skills, etc.). This allows us not only to capture user preferences as norms regulating the behaviour of each agent, but it also allows us to exploit system-wide knowledge of all users' preferences in doing so.

Since this multiagent system has large numbers of users (just considering smart speakers, it is reported that 500 million units were installed as of Q4 2021 \cite{Strategy2021smartspeakerinstalled}), even if there is only partial knowledge of each user's preferences, we can use all available knowledge to make further inferences about their preferences. Inspired by recommender systems, we consider collaborative filtering to be a suitable approach to this end. In recommender systems, collaborative filtering is well-established for predicting items a user may like \cite{Su2009Collaborative}. Given some known user preferences, these methods find similar items the user may like and suggest them to the user. Hence, in a sense, the user profile is completed from predictions with regard to similar items or to users with similar tastes. Following this idea, we can exploit user similarities with regard to their known preferences towards AI technologies to predict unknown preferences. This approach would allow us to construct a full profile of preferences while reducing interactions to a minimum and still capturing the user's real preferences with better success than current approaches.

To motivate this problem, in the following section we examine the particular problem of privacy in AI assistants.
\section{Motivation: Privacy in AI assistants}
\label{sec:motivation}

The advent of the internet has opened a door into a new world of opportunities and faster communication. Nonetheless, this same door has also allowed service providers, organisations, and enterprises to peek into our private lives. Studies show that users are deeply concerned about how their data is being collected online \cite{madden2014perceptions}. For example, 81\% of surveyed users in \cite{madden2014perceptions} feel not very secure sharing private information with trusted persons or organisations through social networks. Yet even though users are concerned about their privacy online, they feel unable to control how their information is being transmitted. Paradoxically, this causes them not to protect their privacy, a behaviour known as privacy cynicism \cite{Hoffmann2016Cynicism}.

Legislators have tried to put regulations in place to protect user privacy, such as those proposed by the European Union \cite{EUGDPR}, the United Kingdom \cite{UKDPA}, or California \cite{CaliforniaCPA}. Among other things, these laws require users to be informed about how their data is being collected and used, as well as requiring their consent for doing so.
However, while these regulations have, for example, increased the number of cookie banners on web pages, in most cases these banners inform or require consent but do not allow users to easily opt-out \cite{Degeling2019wevalue}.
Furthermore, in the cases where users {\em can} opt-out, the banners are designed to guide users towards accepting the privacy policy (for example, by hiding the decline button while highlighting the accept one) \cite{laine2021design}. A user navigating several web pages is thus overwhelmed with a large number and variety of such banners, and evidence shows that these are not effective in capturing the true preferences of the user. As found by Kretschmer et al.~\cite{kretschmer2021cookie}, banners are ignored by users 70\% of the time, while only 10\% of the time are they clicked for privacy reasons. This behaviour is not limited to cookie banners: Obar and Oeldorf-Hirsch~\cite{Obar2018lie} show that most users skip reading privacy policies when signing up to a social network, while those that start reading the policies become overloaded with information and stop reading shortly after. This behaviour can be explained as a result of privacy fatigue, which has been shown to have a stronger effect on users than privacy concerns, and leads them to cynicism \cite{Choi2018Fatigue}.

A particular platform that has recently raised privacy concerns is that of AI assistants. Be it through smart speakers or our mobile phones, we use these assistants to check the weather, set reminders, or even buy things online. However, due to their only very recent adoption, these systems still have numerous security and privacy problems \cite{EduSPASurvey}, which have already led to undesirable situations, such as recording private conversations and sending them to a random contact \cite{GuardianAlexa}. Apart from violating user privacy, these issues threaten user trust in AI assistants, causing users to disable functionalities or to implement coping mechanisms \cite{Abdi2019More}, thus threatening the adoption of the technology. 

In consequence, it is paramount that we ensure AI assistants are secure and align with the privacy expectations of users. This requires a novel method to capture user preferences, one that does not rely on constant and overwhelming user interaction. As an example, imagine that a user asks their AI assistant about the weather, but before completing that task the weather app then asks the user to access their contacts to send them a message recommending that app. The user, who might be in a rush or tired, replies affirmatively, against their true preferences, just to quickly get the weather information. Instead, as an alternative approach, we can examine the pool of users of that app, find some with similar privacy preferences to the user and infer a predicted preference from them. This would at least help to tailor the interaction with the user to the predicted outcome, or even skip the interaction completely, assuming the user had provided the agent consent to decide on these interactions. In this way, we advocate for a technique to capture user preferences that does not require excessive user involvement, and that can be readily deployed in any AI device or service without the need for domain knowledge (i.e., contextual information). This technique should exploit system-wide information (like similarities between users) to build preference predictions.
\section{Related work}
\label{sec:rw}

In our consideration of related work, we review three bodies of literature, namely privacy and AI assistants, normative multiagent systems, and AI ethics.

Firstly, some papers have addressed the problem we have outlined here in the particular domain of privacy and AI assistants. Abdi et al.~\cite{NouraPrivacyNorms2021} conducted a user survey and crowdsourced norms from the commonly agreed opinions of the users. While the approach in this paper would be useful for specifying default preferences, it is not able to tailor them to each individual user. Using that dataset, Zhan et al.~\cite{Xiao2022Model} presented a model to predict privacy norms. Here, norms are predicted using a machine learning model with an accuracy of 70-80\%. Both of these approaches are based on the idea of contextual integrity \cite{nissenbaum2004privacy}, which states that the privacy preferences of users depend largely on the context. For example, in a social network, a user may share a photo with their friends but not with the general public. Here, the receiver of the photo is the context that determines the preferences of the user. Therefore, while Zhan et al.'s work~\cite{Xiao2022Model} seems promising, we argue that a more general approach is needed. Note that constructing predictions based on context requires knowledge of the possible contexts and their connections, and this knowledge is dependent on the domain. For example, the contexts to be considered in a social network will be different from those to be considered in an AI assistant. However, assuming this knowledge will be available is too strong. Instead, as discussed previously, we aim at a technique that can readily be used with any AI device or service without the need for specific knowledge. 

Seeing that both of these works address the problem by resorting to norms, we also consider the normative multiagent system literature to see if any current approaches to constructing norms can be readily adopted for our purposes. Indeed, norms have been long studied in the multiagent systems literature as a means to coordinate agents \cite{BoellaTV06}. In technical terms, there are two main methods for developing norm systems, known as top-down and bottom-up approaches. On the one hand, top-down approaches consider a system-wide perspective and norms are constructed and enacted by a central authority (for example, norm synthesis approaches \cite{MoralesAAMAS2013, morales2015compact}). The aim of these approaches is to exploit system-wide knowledge to build a system-wide norm system. While we also want to exploit system-wide knowledge, we differ in our goals, as we are concerned with developing individual norm systems for each agent. On the other hand, bottom-up approaches construct norms at an agent level based on the agent's experience and view of the environment (for example, norm emergence approaches \cite{shoham1997emergence, SavarimuthuAAMAS07, sugawara2011emergence}). Here, while norms are built at an agent level, they are only created considering agent level knowledge which, for our purposes, would require unwanted user interaction. Instead, the method we envision has elements of both top-down and bottom-up approaches, in that a norm system would be built individually and independently for each agent but considering system-wide knowledge of all agents' preferences to do so. Another aspect of differentiation between approaches is the time at which norm systems are constructed. On the one hand, off-line approaches like those in \cite{ShohamT95, agotnes2007logic}, build norm systems at a point in time, considering the information they have at that point. In contrast, given that user preferences may change over time and new situations will arise, our aim is for an online approach, like those in \cite{MoralesAAMAS2013, morales2015compact}, which are able to construct norms during execution of the multiagent system or, in our case, during the life of the AI device or service in question.

Another focus of the literature is producing norm systems that satisfy desirable properties. There has been vast research on building norm systems satisfying properties such as: conflict-avoidance \cite{Kollingbaum2007, VasconcelosKN09} (ensuring that the resulting norms do not conflict with each other); minimality \cite{fitoussi00choosing, MoralesAAMAS2014} (avoiding over-regulation); compactness \cite{morales2015compact} (producing small norm systems); and many others. While these formal properties are important, our main focus is the alignment between user preferences and norms (regardless of the properties satisfied). In this sense, the approach to solving our problem should be closer to that of the field of AI Ethics. Recent work here has studied the engineering of moral value-aligned norm systems. For example, the work of Serramia et al.~\cite{serramia2018aamas, serramia2021JAAMAS} composes sets of norms considering (and maximising) their alignment with moral values and known moral value preferences. Sierra et al.~\cite{sierra2019value} propose a formal approach to finding value-aligned norms based on guiding the multiagent system to those states that promote the desirable moral values. Finally, Montes and Sierra~\cite{montes2021aamas} tweak parameters in norms to maximise their promotion of moral values (such as equality or fairness). Note that moral values are the criteria by which we distinguish between good and bad behaviour \cite{charisi2017moral, Dignum2017responsible}. Therefore, ensuring AI works as expected by the user is in turn making AI aligned with the user's morality. For example, applying our approach to the AI devices of a user that highly values privacy will result in those devices respecting the user's privacy. Therefore, while work in the area of AI and ethics could be used to address the goals of this paper, they require knowing the moral value preferences beforehand, which is also too strong an assumption. Furthermore, all these approaches are off-line, which are unable to adapt to changing environments.

To conclude, while contextual integrity or morality are highly relevant to assert a user's preferences, we cannot assume explicit knowledge of either of these two aspects. Hence, our aim should be a new technique that does not require this knowledge.
\section{Our proposal: Capturing user preferences as norms}
\label{sec:proposal}

In this section we discuss several ideas to address the shortcomings we have identified in the literature. Firstly, we consider the possibility of exploiting user similarities in order to predict user preferences. Secondly, we discuss applying this idea to the task of specifying a collaborative filtering method for predicting a user's preference norms.

\subsection{Exploiting user similarities}

A common trait of AI devices and services is their large pool of users, and the vast amount of available information about them. As a result, it might not be necessary to ask a user for all their preferences, and we might instead predict them. Note that we have partial knowledge over the preferences of each user. Individually, these preferences can be seen in bounded numerical form\footnote{For numeral preferences, since the number of users and preferences is finite, there will always be an upper and lower bound. For ordinal preferences, we can define a transformation function that transforms an order into numerical preferences in a bounded interval.}. Thus, collectively these preferences are defined in a multi-dimensional real space (where the number of dimensions is the number of elements over which preferences are defined, i.e. the actions that AI devices can perform). This multi-dimensional space represents all possible preferences over all possible behaviours of the AI device in question. Then, the preferences of each individual user of the AI device are a point in this space (where each position of the point refers to a preference with regard to some action the AI device can perform). Since not all preferences are known, each user's point has many gaps (i.e. a gap in each position of the point representing an unknown preference). But even just considering the known positions in the point, we can establish distances between user preference profiles. For example, considering the real distance for known preferences and maximum distance for unknown ones. The overall distance between two users tells us how similar or dissimilar users are with regard to their preferences. Those with small distances will have very little divergence of preferences and therefore can be considered similar with regard to them. 

Note that user preferences are not independent; for example, in the case of privacy preferences users might want to protect their privacy, leading to preferences against sharing data, or they might not care about their privacy, leading to preferences in favour of sharing data. Therefore, it is highly probable that, if users are similar with regard to their known preferences, they will also be similar with regard to unknown preferences. Therefore, even if we only have partial preferences for most users, we can still predict the missing preferences by inferring them from similar users. While this idea has been commonly used in areas like recommender systems, to the best of our knowledge it is a novel idea in the context of setting preferences about how an AI agent should behave. Importantly, this approach would not require knowledge of what the preferences mean, or information about contexts or the morality of the user which, as we have seen in Section \ref{sec:rw}, are common assumptions in the literature. Instead, this approach can readily be used in any application where we know partial preferences for each of the users, without the need for extra information. In fact, in this case, knowledge about the context and morality of decisions is considered implicitly, as similar users might have similar morality and views on contextual integrity.

\subsection{Collaborative filtering to predict norms}

When it comes to specifying how AI agents should behave, we resort to norms. Firstly, norms serve as a concise way to define preferences in complex domains with a multitude of contexts, roles, etc. As new norms are created, we can combine them into more general norms. We are also able to detect inconsistencies (i.e.\ the newly created norm may contradict already established norms) and resolve them (for example, by abolishing the older norm). Secondly, agents may represent several humans with different preferences, in the particular case of AI assistants it is common for whole families to use a single smart speaker. Hence, norms can be a useful means to reason about multiple users' preferences and reach a consensus set of norms representing them all. Thirdly, considering the ecosystem of AI devices or services as a multiagent system, norms could help create an expectation or provide guidance for behaviour to other devices or services in this ecosystem. For example, in the case of AI assistants, if a user has several norms prohibiting the device from sharing information during night-time, other services requiring information might wait until daytime to request it. Finally, norms not only serve as a means to regulate agent behaviour but are also useful to {\em explain} agent behaviour as they are natural for humans. Thus, we can use the user's norms to construct explanations of agent behaviour that are easily understood by them, if need be. Importantly, since this process does not require explicit knowledge of context, roles, etc. it can adapt to different definitions of norms, as simple or as complex as required by the application domain.

Adopting the approach of the previous subsection, we are able to predict preferences of a user with regard to how their AI devices should behave. Note, however, that here user preferences strictly determine how the agent should act. If a user of an AI assistant sets their preferences to not share any data, it should imply that the AI assistant will not share any data. Therefore, depending on the predicted preferences, we can directly infer a norm regulating the behaviour of the AI agent. A strong negative predicted preference toward some action implies the agent should be prohibited from performing that action. Conversely, a strong preference in favour of the action, should imply the agent is obliged to perform it. Thus, in essence, when performing collaborative filtering we are not only predicting user preferences, we are also predicting norms.

Technically speaking, predicting norms is a two-strep process: first we predict the numerical preference, and then we feed it into a function that transforms the numerical preference into a norm.
Note, however, that only clearly positioned preferences must lead to a norm. Neutral or close to neutral preferences are not strong enough to infer a norm, so in these cases the function should result in no norms being constructed. 
Furthermore, even in cases where we have clear predictions, we must consider prediction confidence to minimise the chance of constructing incorrect norms. Prediction confidence assesses the quality of the knowledge used to make the prediction. This can be calculated from both the similarity measured between the target user and its similar users, and from the amount of known preferences they have in common. Therefore, we can fix a threshold of necessary confidence to construct norms. Secondly, we may also consider context sensitivity (as has been considered previously when constructing privacy norms with regard to information transmissions in online social networks \cite{Criado2015Implicit}). In our case, sensitivity can be inferred from the pool of users; for example, contexts that are usually part of norms forbidding interaction might be sensitive. In sensitive contexts, it might be better to interact directly with the user instead of constructing predicted norms.

Note that whenever interaction is needed (e.g., sensitive contexts), we can still tailor the interaction using our predictions. Hence, we argue that with this less invasive and more meaningful interaction, users will pay attention to consent requests and provide truthful answers. These truthful answers and norms, stemming from high confidence predictions, will ensure that the user's real preferences guide the behaviour of agents in the multiagent system.

Finally, it is worth mentioning that norm predictions are also useful to detect preference changes. As time progresses, we gather more knowledge on all users, so that when a preference norm has been in place for some time, we can test if it still holds by re-predicting it. If the predicted preferences remain the same, we maintain the norm; whereas if they have changed we can update the norm directly or interact with the user if we lack prediction confidence or the context is sensitive.
\section{Conclusions}
\label{sec:conclusions}

Current methods to capture user preferences are ineffective and need to be improved. This is particularly crucial for AI technologies, which take decisions independently and therefore must understand user preferences to behave as users would expect. Methods relying on constant interaction and questioning annoy users and prevent them from achieving their goals when using these technologies. Furthermore, they fail at obtaining real answers, as usually users answer automatically just to swiftly deal with questions. Given the large amount of information already available from the large pool of users using the same devices and services, we advocate for minimising interaction, and for predicting preferences from knowledge of other users. This will decrease the amount of annoying interruptions, and therefore give more importance to questions. Users are therefore more likely to give truthful answers, as the smaller number of interactions might not be perceived as constant disturbance but as warnings. By completing the preferences with predictions, we should be able to obtain a complete set of preferences reasonably close to reality, or at least much better than those obtained with current methods. Note, too, that for critical preferences in which interactive questioning is required (e.g., when dealing with highly sensitive data), preference prediction could be used to rephrase the question to ask, for example for confirmation of the predicted preferences.
\bibliographystyle{splncs04}
\bibliography{Bibliography}
\end{document}